\documentstyle{article}
\newcommand{\bibi}{\bibitem}                                                  
                                             
\newcommand{\prl}{\it Phys. Rev. Lett.}                                       
                                           
\newcommand{\prb}{\it Phys. Rev. B}

\newcommand{\half}{\frac {1}{2}}                                             
                                             
\newcommand{\beq}{\begin{equation}}                                           
\newcommand{\eeq}{\end{equation}\noindent}                                  
\newcommand{\beqr}{\begin{eqnarray}}                                          
\newcommand{\eeqr}{\end{eqnarray}\noindent}                                   
                                                      
\newcommand{\vr}{{\bf r}}                                                     
\newcommand{\vk}{{\bf k}}                                                     
                                                     
\newcommand{\vq}{{\bf q}}                                                     
\newcommand{\cb}{{\rm b}}                                                     
\newcommand{\ch}{{\rm h}}                                                     
\newcommand{\cbd}{{\cb}^{\dag}} 
\newcommand{\chd}{{\ch}^{\dag}}

\newcommand{\cd}{c^{\dag}} 
\newcommand{\bd}{b^{\dag}}                                                    
\newcommand{\hd}{h^{\dag}}

\newcommand{\noin}{\noindent}                                                 
\begin{document} 
\title{Spin-Charge Separation and Kinetic Energy in the t-J Model} 
\author{Sanjoy K. Sarker \\                                                   
Department of Physics and Astronomy \\                                        
The University of Alabama, Tuscaloosa, AL 35487 \\}                           
\maketitle    
\begin{abstract}

The effect of spin-charge separation on the kinetic energy of 
the two-dimensional $t$-$J$ model is examined. Using a sum rule,
we derive an exact expression for the lowest possible KE
($E_{bound}$) for any state without doubly occupied sites. 
The kinetic energies of the relevant slave-boson and 
Schwinger-boson mean-field states are found to be considerably
larger than $E_{bound}$. The MF states 
exhibit complete spin-charge separation, and  form the basis
of a number of microscopic theories. An examination of the
momentum distribution reveals that the large increse in KE 
of the MF states is due to excessive depletion of electrons
from the bottom of the band (Schwinger boson states) and holes
from the top of the band (slave boson states). To see whether
the excess KE is simply due to the poor treatment 
of the local constraints, we solve the constraint problem 
analytically for the Schwinger boson MF states in the  
$J = 0$ limit. This restores gauge invariance, incorrectly
violated in the MF theories. The resulting state is a 
generalization of the Hartree-Fock state in the Hubbard model, 
but one that includes spin-wave excitations, removing a 
deficiency of the simple HF theory. Even after the 
constraints are imposed correctly, the MF kinetic energy
is found to be much larger than $E_{bound}$.
These results support the notion, advanced in earlier
papers, that spin-charge separation in the MF state costs 
excessive kinetic energy, and makes the state unstable toward
recombination processes which lead to superconductivity
in $d = 2$ and a Fermi liquid state in higher dimensions.          

   
\end{abstract}                                                                
 


\centerline{I. INTRODUCTION}
\bigskip

The possiblity of spin-charge separation in large-$U$ Hubbard model 
and the equivalent $t$-$J$ model is of considerable interest because 
of its relevance to high-$T_c$ superconductivity \cite{and1}. Such a
behavior is seen in $d = 1$, where the metallic state is characterized 
by separate spin and charge excitations, not electron-like 
quasiparticles \cite{and2}. The metallic state evolves out of
the insulating state at half filling, which has no magnetic order. 
By contrast, for $d \geq 2$, the ground 
state is antiferromagnetically ordered at half filling. 
Therefore, if the 2-d Hubbard model were to describe the cuprates, its 
low-energy behavior would be governed by several distinct fixed 
points: an AF insulator; a metallic, presumably spin-charge separated 
state without magnetic order; possibly a superconducting state; and 
possibly even a Fermi liquid at lower electron densities. 

A number of microscopic theories are based on the slave-boson (SLB) 
\cite{zou,nag} and the Schwinger-boson (SWB)\cite{jay} representations 
of the 2-d $t$-$J$ model in which the electron field is decomposed into 
a spin (spinon) and a charge (holon) component, subject to a local 
constraint. Each theory is centered on a mean-field state which is
supposed to take care of short-range or high-energy processes,
leaving behind weak residual interactions. The MF state is
characterized by independent spinon and holon excitations, and 
thus exhibits complete spin-charge separation. The results have been 
varied, but not too satisfactory. That none of the MF theories 
succeed suggests that there is a generic reason for the failure -- 
some important bit of physics is missing in all of them. Identifying
it and understanding the energetics behind the failure
is of great interest.

One obvious source of error, to be considered later, is the local 
constraint, which is treated on the average in all MF theories. 
We have argued previously on continuity grounds
that there is another, and more fundamental, reason for the 
failure \cite{sar1,sar2}. The MF states exhibit maximum spin-charge 
separation in all dimensions and at all densities. This can not 
be true, since in high dimensions (e.g., in $d = 3$), one expects 
a Fermi liquid state to emerge. In $d = 1$, strong spin-charge 
separation occurs because electrons of opposite spins can not pass 
each other, and thus are localized as \lq\lq spins" or \lq\lq local 
moments". This costs extra kinetic energy. In higher dimensions, 
electrons can avoid double occupancy by going around each
other and delocalize, and thereby reduce the kinetic
energy. Therefore, spin-charge separation should weaken
with increasing dimensionality and decreasing electron density. 
MF states fail because they show too much spin-charge separation 
which costs excessive kinetic energy. They would be unstable
toward spin-charge recombination into physical electrons, the latter 
would try to form a Fermi liquid.
Recombination will help destroy magnetic order because, as the
electrons delocalize, the size of the moment is reduced.
It will be more effective at higher dimensions and far
from half filling since more pathways are available, which would 
explain the emergence of the Fermi liquid state in $d = 3$.

Spin-charge recombination is a collective process which appears 
through the kinetic energy term. It has been analyzed in $d = 2$
using RPA in the Schwinger-boson representation.
The electron is a \lq\lq collective" excitation. Recombination
is strong enough to destroy magnetic order, 
but the normal state remains weakly spin-charge separated \cite{sar1}.
Additional spin-charge recombination, 
in which a pair of holons combine with a pre-paired spin singlets
then leads to $d$-wave superconductivity \cite{sar2}. In this
picture the destruction of magnetic LRO, the appearance of 
superconductivity and the emergence of a Fermi liquid state
at high dimensions are all caused by spin-charge  
recombination, as the system evolves away from half filling.

In this paper we present two exact results which allows us
to get more insight into the energetics of spin-charge
separation. Specifically, we examine the hypothesis that
MF states have excessive kinetic energy. 
This issue is quite subtle. How does one know how much KE is 
too much? With increasing $U$, kinetic energy of the actual
ground state will increase, as electrons are excited out
of the Fermi sea. As $U \rightarrow \infty$, there is no 
double occupancy, and we expect KE to be quite large 
(compared with the noninteracting case, $E_{free}$), even if 
the state is a Fermi liquid. By using a sum rule we determine 
the lowest possible kinetic energy ($E_{bound}$) for any state of
the $t$-$J$ model, i.e., for any state in which there is no 
double occupancy. As expected, we find that $E_{bound} >> 
E_{free}$.  Interestingly, near half filling, the 
Schwinger-boson state has a lower KE than the slave-boson state.
However, as we shall see, the kinetic energy of either state is much 
larger than $E_{bound}$. An examination of the momentum 
distribution reveals that excess KE is due to excessive
depletion of single particle states near the band edges -- a 
situation that is  quite likely to give rise to strong fluctuations. 

The second issue involves the largely unsolved problem of the 
constraints. Treating them on the average would be valid if
fluctuations were small enough to be treated perturbatively.
However, a perturbation theory is not likely to converge, 
and may make matters worse. The constraints
imply a local gauge symmetry which, according to Elitzur's 
theorem \cite{eli}, can not be spontaneously broken. MF theories 
break this symmetry so that elementary excitations, i.e.,
spinons and holons, are not gauge invariant and therefore not
observable. The true ground state and elementary excitations
are of course gauge invariant.  Since gauge invariance can not be 
restored perturbatively it is hard to judge the correctness of
the theory, even at a qualitative level.  

One way to avoid the constraint problem is to consider the 
Hartree-Fock (HF) states of the Hubbard model \cite{sar3}, 
in which constraints are treated exactly. These states  
are related to the Schwinger boson MF states. But the HF
theory has its own shortcomings. The spin-wave excitations
are missing at the HF level, so that low-energy behavior
is not described correctly. For example, magnetic order
persists all the way upto a $T_c \sim U$ in all dimension,
violating the Marmin-Wagner theorem.
On the other hand, for the Heisenberg 
model, the Schwinger boson MF theory is qualitatively similar to 
the spin-wave theory, which is based on the Holstein-Primakoff (HP)
representation and is gauge invariant \cite{aro,sar5}. 
But HP theory applies only to half-filling, there is no charge 
excitations. 

Here we solve these problems in the $U = \infty$ limit. This limit 
is of interest because the mapping onto the $t$-$J$ model is exact, 
so a direct comparison between SWB and HF results can be made. 
By integrating out the constraints in the Schwinger boson theory, 
we obtain a generalized HP representation which is valid away from 
half filling. This, in turn, leads to a generalized Hartree-Fock 
Hamiltonian, which include both charge and spin excitations. 
The procedure restores gauge invariance, and the inclusion of 
the spin-waves takes care of the deficiencies of the HF theory. 
We find that even after the constraints are taken into account,
the nature of the state is not changed, nor is the degree
of spin-charge separation. Kinetic energy is reduced, but
continues to be excessive compared with $E_{bound}$, which   
supports our hypothesis.
 
\bigskip

\centerline{II. THE MODEL AND SPIN-CHARGE SEPARATION}

\bigskip

We start with a discussion of the meaning of spin-charge separation.
The Hubbard model is characterized by nearest-neighbor hopping $t$, 
and on-site repulsion $U$. For large $U$, doubly occupied sites can
be projected out, which leads to the $t$-$J$ Hamiltonian:
\beq H = - t\sum _{ij}(\cd _{i\sigma}c_{j\sigma} + h.c.) +
J \sum _{ij}({\bf S}_i.{\bf S}_j - \frac{1}{4}n_in_j), \eeq
where the summation is over nearest neighbors. The second term
describes superexchange, with $J = 4t^2/U$. The first term describes 
hopping, and has the same form as that in the Hubbard model. 
However, $\cd _{i\sigma}$ creates a \lq\lq projected electron",
i.e., an electron of spin $\sigma$, only if the
site $i$ is unoccupied. At half-filling, one needs to consider only
the exchange term. 

In $d = 1$, the Hubbard model has been diagonalized by 
Lieb and Wu \cite{lie}, using the Bethe Ansatz. The occurrence of 
spin-charge separation can be seen most clearly in the large $U$ 
limit where, as shown by Ogata and Shiba \cite{oga}, 
the ground-state wave 
function for $N_e$ electrons can be approximately written as:
\beq \psi = \psi _{charge}\psi _{spin}. \eeq 
Here $\psi _{charge}(x_1,x_2,..)$ is a determinantal wave function 
for $N_e$ {\em spinless} fermions, where $x_i$ are the electron
coordinates, irrespective of spin. And $\psi _{spin}$ is the Bethe's 
solution for the Heisenberg model for $N_e$ spins (with the
holes removed). The wave function at smaller $U$ is more
complicated, but is continuously connected to the large $U$
solution in the renormalization group sense. Of course, the 
low-lying excited states also similarly decompose -- i.e., the
system is characterized by independent spin and charge 
excitations. A good account of the meaning of Eq. (2) and its 
consequences is given by Anderson \cite{and2}.

Usually spin-charge separation is associated with the Luttinger
liquid state, which is a metal and has no magnetic order.
However, Eq. (2) clearly suggests that the phenomenon is more 
general. Any wave function which has the form of (2), or is 
continuously connected to such a wave functions, 
exhibits spin-charge separation. Consider the following
cases.

a) Half-filled case: If we continue the
Lieb-Wu wave functions to half-filling, the form (2) will 
clearly be preserved. The (spinless) charge band becomes filled, 
and the system turns into an insulator. A moment's reflection shows 
that such spin-charge separation occurs at half filling
in all dimensions. The reduction of the large-U 
Hubbard model to the Heisenberg model is precisely 
due to spin-charge decoupling. There is exactly one charge per site, 
which can be viewed as a filled band of spinless fermions.
For $d \geq 2$, the insulating state is magnetically ordered. 

b) Mean-Field States: Similarly, mean-field wave functions for the 
$t$-$J$ model also
exhibits spin-charge separation. In this case the electron operator is 
represented as:
$$\cd _{i\sigma} = \bd _{i\sigma}h_i, $$
where $\bd _{i\sigma}$ creates a spinon and $h_i$ creates a holon,
subject to the constraint
$$ \bd _{i\uparrow}b_{i\uparrow} + \bd _{i\downarrow}b_{i\downarrow}
+ \hd _ih_i  = 1. $$
The spin-operators are then represented as: $S^{+}_i = 
\bd _{i\uparrow}b_{i\downarrow}$, and $S^{z}_i = \frac{1}{2} 
(\bd _{i\uparrow}b_{i\uparrow} - \bd _{i\downarrow}b_{i\downarrow})$.
In the Schwinger boson representation, the spinon is a boson and holon,
a fermion. The opposite is true in the slave boson representation.
In either case, constraints are treated on the average, and a MF
approximation leads to the Hamiltonian:
\beq H_{MF} = H_{spinon} + H_{holon}. \eeq
Note that all eigenstates of $H_{MF}$ are of the form of Eq.(2).
This is also true in the presence of arbitrary spinon-spinon and 
holon-holon interactions. 

In general bosons condense, leading to long-range order (LRO). 
In the Schwinger-boson case, the ordering is magnetic.
The mean-field theory gives quite a good description of the Neel 
state at half filling \cite{aro}.  Away from half filling, the AF 
insulator evolves continuously into a metallic state with a spiral
LRO, and eventually to a ferromagnetic metal at 
small $J/t$ \cite{jay}. In the metallic region, $\psi _{charge}$  
in the SWB theory is again a spinless fermion determinant.

c) Hartree-Fock states: The magnetically ordered states obtained by 
the Hartree-Fock treatment of the Hubbard model \cite{sar3} constitute 
a third example of spin-charge separation. The 
HF decomposion of the $U$ term leads to the quadratic Hamiltonian
\beq H_{HF} = H_0 - 2U\sum _i {\bf m}_i.{\bf S}_i,  \eeq
where $H_0$ is the hopping part and ${\bf m}_i = <{\bf S}_i>$ is the 
local magnetization, which is determined self consistently.
To describe spiral states, it is convenient to break the spin 
rotational symmetry in the $xy$ plane and choose $<S^+_i> = me^{iQ.r}$
as the order parameter, where $Q$ is the spiral wavevector which 
continuously connects the AF state ($Q = (\pi,\pi)$) at half-filling,
to the ferromagnetic state ($Q = 0$) which occurs at large $U\delta/t$.
The theory has been described in detail elsewhere \cite{sar3}. 
The HF approximation leads to two magnetic bands separated by a gap of 
size $2Um$. At large $U$, low-energy states are confined to the lower band. 
However, there is no spin-degeneracy; each one-particle state can 
accomodate one electron, not two. Therefore, low-energy wave functions are 
one-component (i.e., spinless) determinantal wave functions, 
similar to $\psi _{charge}$ in Eq. (2). The HF states are spin-charge 
separated because they are closely related to the corresponding
states of the Schwinger boson MF theory.  

To see this, consider $H_{HF}$ for $Um >> t$. Then, the 
spin of an electron will locally be parallel to the local moment 
${\bf m}_i$. Let $f^{\dag}_i$ and $g^{\dag}_i$ 
create electrons with spin parallel and antiparallel to ${\bf m_i}$,
respectively. The bare electron operators, which are quantized along the
$z$ direction, can be expanded as:
\beq a_{i\sigma} = u_{i\sigma}f_{i\sigma} + 
sgn(\sigma)u^*_{i,-\sigma}g_{i\sigma}, \eeq
where  $\sigma = 1$ for $\uparrow$ and $- 1$ for $\downarrow$.
The complex numbers $u_{i\sigma}$ depend on the direction of ${\bf m}_i$,
and satisfy the normalization condition
\beq \sum _{\sigma}u^*_{i\sigma}u_{i\sigma} = 1, \eeq
Suppose, ${\bf m}_i = (m,\theta,\phi)$, then a simple representation is: 
$$u_{i\uparrow} = \cos (\theta _i/2), ~~~~~ 
u_{i\downarrow} = \sin (\theta _i/2)e^{i\phi _i}. $$
Note that the second term in $H_{HF}$ is 
diagonal in the $f,g$ representation:
$$-Um \sum _i(f^{\dag}_if_i - g^{\dag}_ig_i). $$
For $U \rightarrow \infty$, we can neglect the $g$ terms which 
constitute the upper band. The similarity with the Schwinger boson
representation is obvious, with $u_{i\sigma}$ playing the role of 
$b_{i\sigma}$ and $f_i$ the charge part of the electron operator.
The holon operator is obtained by 
a particle-hole transformation: $f_i \rightarrow h^{\dag}_i$.
Eq. (6) now plays the role of the constraints. The difference is that 
$b$'s are operators, where as $u$'s are c-numbers. 
However, in the MF approximation, $b$'s condense, and we can take 
$u_{i\sigma} \approx <b_{i\sigma}>. $
For $U = \infty$, the ground state
is a ferromagnetic metal away from half filling. Then ${\bf m}_i =
{\bf m}$, (and hence $u_{i\sigma} = u_{\sigma}$), is independent of $i$. 
The HF wave function, written in terms 
of the original electron coordinates, (those created by bare operators
$a_{i\sigma}$),
spin-charge separates, with $\psi _{charge}$ again given by 
the one component Slater determinant, and is independent of ${\bf m}$.
$\psi _{spin}$ describes the condensed part and is just a constant. 

For finite but large $U$, there will be small admixture of $g$'s via 
the hopping term which can be eliminated perturbatively, 
to obtain contribution of order $J$ --- the exchange contribution.
However, as in the $t$-$J$ model, such corrections are due to
spin-spin interaction and thus does not change our conclusion about
spin-charge separation. There are of course some important differences
between the two theories. In the SWB MF theory, constraints are treated
on the average, whereas in HF theory they are treated exactly. 
Similarly, in the HF theory, the low energy spin excitations are
absent, and hence physics of the singlets is not included.

\medskip

\noin {\em {Spin-Charge Separation and Local Moments}}

\medskip

At half-filling, it is natural to associate spin-charge separation
with the formation of local moments. In $d = 1$, evidently
the local moment picture persists away from half-filling since 
$\psi _{spin}$ continues to be the wave function of a \lq\lq squeezed" 
Heisenberg antiferromagnet \cite{oga,and2}.
At large $U$, two electrons of opposite spin can
not go past each other. Hence, electrons are localized within a
region of size $(1 - \delta)^{-1}$, and the size of the moment is 
approximately $(1 - \delta)/2$. On the other hand, if they can
delocalize there will be no moment, nor will there be spin-charge
separation. Therefore we can associate spin-charge separation with
\lq\lq localized" moments. Evidently this picture is consistent 
with Schwinger boson MF states, as well as HF states.  
Note that spin-charge separation is associated with the amplitude,
and does not require the existence of LRO (phase coherence).
The latter disappears at any
finite $T$ in $d = 2$ (Mermin-Wagner theorem). But the
wave-function remains spin-charge separated since, at the 
MF level, the moment vanishes only at $T \sim U$. Because moment
formation involves a type of localization, it would cost kinetic energy.
Interestingly, from this point of view a transition from a Luttinger 
liquid to a Fermi liquid can be considered a form of delocalization 
transition.

\bigskip

\centerline{III. SPECTRAL SUM-RULE AND LOWER BOUND ON KINETIC ENERGY}

\bigskip

In this section we obtain an exact expression for the lowest possible 
KE in the $t$-$J$ model. It is useful to express 
the hopping part of the $t$-$J$ Hamiltonian in the momentum space
\beq H_{hop}  
= \sum_{\vk,\sigma} \epsilon(\vk)\cd_{\vk\sigma}c_{\vk\sigma}, \eeq
where $\epsilon (\vk) = - 2t(\cos k_x + \cos k_y)$ is the free-electron
hopping energy, and $\cd_{\vk\sigma}$ is the projected electron operator
in the momentum space.
The average hopping energy (per site) is then given by
\beq E_{kin} = 
\frac{1}{N}\sum _{\vk}\epsilon (\vk)n_c(\vk), \eeq
where $n_c(\vk) = \sum _{\sigma}n_{c\sigma}(\vk)$ 
is the (total) electron momentum distribution, with 
$n_{c\sigma}(\vk)  = <\cd_{\vk\sigma}c_{\vk\sigma}>$.
Eq.(8) is exact. Hence any errors in $E_{kin}$ are due to 
approximations
used in calculating $n_c(\vk)$. The kinetic energy of the Hubbard
model is given by the same formula, except that the corresponding
momentum distribution differs from $n_c(\vk)$ by terms of order 
$J$, and $J\delta$.  Here, we focus on the kinetic energy as 
defined above. Of course, for $U = \infty$ (i.e., $J  = 0$), 
the mapping is exact.

Usually one is interested in the singularities of $n_c(\vk)$ near the 
Fermi surface. However, in the present case, we need to know $n_c(\vk)$ 
for all $\vk$.  In order to minimize the kinetic energy, the system
will try to make $n_c(\vk)$ as large as possible near the
bottom of the noninteracting band $\epsilon (\vk)$, and as small as 
possible near the top. Hence states with $\vk$ near the band edges
are quite important. This has serious implications for the mean-field 
states. The assumption underlying the microscopic theories is  
that high energy processes, i.e., those involving electrons with $\vk$
far from the Fermi sea, are accurately taken care of at the mean-field 
level, leaving behind residual interactions which describes low-energy 
physics. Therefore, if a MF state does not yield accurate values
of $n_c(\vk)$ near the band edges, it is likely to be unstable 
toward fluctuations which transfer electrons from the top to the bottom
of the band, and thereby lower the kinetic energy considerably.

\bigskip

\noin A. {Spectral Sum Rule:}

\medskip 

Now, the noninteracting state has the the lowest KE, 
with $n_c(\vk) = 2$ below the Fermi level $\epsilon _F$.
However, this is not a good reference state since any interaction 
would increase the kinetic energy by exciting electrons out of
the Fermi sea and, as we see below, at large $U$, a substantial part of
of the increase comes from avoiding double occupancy, which may
have nothing to do with spin-charge separation. Fortunately,  
we can determine the lowest possible kinetic energy in the 
$t$-$J$ model. Since the latter does not have double occupancy, 
the corresponding increase in KE is accounted for at the outset.

Formally, we have
\beq n_{c\sigma}(\vk) = \int\frac{dz}{2\pi}A_{c\sigma}(\vk,z)f(z), 
\eeq
where $A_{c\sigma}(\vk,\omega)$ is the spectral
function for projected electrons which obeys
a sum rule \cite{sumr}.
Here we give a derivation which is useful
for this paper. Let us define the momentum distribution
for physical holes (not to be confused with holons):
$p_{c\sigma}(\vk) \equiv <c_{\vk\sigma}\cd_{\vk\sigma}>$.
This is obtained from 
\beq p_{c\sigma}(\vk) = 
\int\frac{dz}{2\pi}A_{c\sigma}(\vk,z)(1 - f(z)). \eeq
Hence, total spectral weight for spin $\sigma$ is given by
\beq \int \frac{dz}{2\pi} A_{c\sigma}(\vk,z)  
= n_{c\sigma}(\vk) + p_{c\sigma}(\vk) 
= <\lbrace c_{\vk\sigma},\cd_{\vk\sigma}\rbrace>. \eeq
For usual fermions operators, the anticommutator equals
unity, giving the usual spectral sum rule. This
is not true for projected electrons. To evaluate
the right hand side, we use the representation of the electron
operator in terms of spinon and holon operators which, in momentum
space, gives
 $$\cd_{\vk\sigma} = 
N^{-\half} \sum _{\vq}\bd_{\vk+\vq,\sigma}h_{\vq}.$$
Then the average value of the anticommutator is given by
\beq <\lbrace c_{\vk\sigma},\cd_{\vk\sigma}\rbrace> =
\frac{1}{N}\sum _{i}(<\bd_{i\sigma}b_{i\sigma}> + <\hd_ih_i>)
= n_{b\sigma} + \delta, \eeq 
where $n_{b\sigma}$ is the average number spinons (per site) 
of spin $\sigma$, and $\delta$ is the average number holons per
site. It is useful to separate into a charge and a spin part. 
Let $n_{b\sigma} = n_b/2 + \sigma m$, where 
$m = (n_{b\uparrow} - n_{b\downarrow})/2$ is the average
magnetization, and $n_b = \sum _{\sigma} n_{b\sigma} = 1 - \delta$ 
(from the constraint condition). Then we have
\beq n_{c\sigma}(\vk) + p_{c\sigma}(\vk) = 
\half (1 + \delta) + \sigma m. \eeq
We stress that these sum rules are exact. Since the right hand 
side depends on the spinon and holon densities the sum rules  
are valid in either representation, and in all dimensions, 
irrespective of whether the local constraint is imposed exactly or 
on the average. 
Summing over $\sigma$ we obtain the total electron spectral weight:
\beq n_c(\vk) + p_c(\vk) = (1 + \delta). \eeq
The remaining $(1 - \delta)$ resides in the upper Hubbard band.
Note that, in contrast to bare electrons, the spectral weight (Eq. 13)
depends on charge and spin densities. 

The sum rule, Eq. (14), implies that each momentum state 
$\vk$ can accomodate at most $1 + \delta$ electrons
(compared with 2 for a noninteracting system).
It follows that, at half filling, in order to accomodate
all the electrons, one must have
$n_c(\vk) = 1$ for all $\vk$, which is the correct result.
In the Schwinger boson MF theory this can be viewed as a filled
holon band. The kinetic energy per electron is maximum ($ = 0$), 
as compared with $\sim -t$, for free electrons.

\medskip

\noin B. {\em Lower Bound for the Kinetic Energy:}

\medskip

Away from half filling, the exact $n_c(\vk)$ is not known. But it
is clear that states with low kinetic energy will have $n_c(\vk)$       
large near the bottom of the band and small near the top. Indeed,
the lowest possible kinetic energy for the model
is obtained by filling up a {\em modified Fermi sea} with exactly 
$n_c(\vk) = (1+\delta)$, upto a new Fermi level $\epsilon _{F1}$.
Above $\epsilon _{F1}$, $n_c(\vk) = 0$.
This gives an exact lower bound for the kinetic energy per site.
\beq E_{bound} = \frac{1+\delta}{N}\sum _{\vk}\epsilon (\vk) 
\theta(\epsilon _{F1} - \epsilon (\vk)), \eeq
where $\epsilon _{F1}$ is determined from
$$ n_c = \frac{1+\delta}{N}\sum _{\vk}  
\theta(\epsilon _{F1} - \epsilon (\vk)). $$
where $n_c = 1 - \delta$. The modified Fermi sea is obtained 
from the noninteracting one by exciting $1 - \delta$ electrons
from each state below the (noninteracting) Fermi level $\epsilon _F$ 
to a state
above. Therefore $\epsilon _{F1} > \epsilon_F$, and the 
kinetic energy is correspondingly larger (see Fig. 1).
This cost in kinetic energy is due to the removal double occupancy.
In other words, $E_{bound}$ is the lowest possible kinetic
energy for any state in which double occupancy is forbidden.

Apart from the prefactors, the expression for $E_{bound}$ at
a density $n_c = 1-\delta$ is identical to that of a noninteratcing
system at a  higher density $n^{\prime} = 2n_c/(1 + \delta)$:
\beq E_{bound}(n_c) = 
\frac{1 + \delta}{2}E_{free}(\frac{2n_c}{1 + \delta}).\eeq
The volume of the modified Fermi sea is $2/(1 + \delta)$ times the free
electron Fermi sea.        
 
What sort of a state corresponds to $E_{bound}$? Since
it evidently resembles a free-fermion state with
a sharp Fermi surface, let us
consider the following wave function:
$$ |\Psi>  = {\cal Z}\Pi_{k<k_F,\sigma}~ \cd _{\vk\sigma}|vac>,$$
where $\cal Z$ is a normalization constant. Since 
$c_{\vk\sigma}$ represents a projected electron, it is easy to 
see this is
the Gutzwiller wavefunction in the infinite-$U$ limit. 
Now, the projected electron operators for arbitrary spins
satisfy the following commutation rules   
$$ \lbrace c_{\vk\sigma},\cd _{\vk^{\prime},\sigma^{\prime}}\rbrace
= \frac{1}{N}\sum _ie^{i(\vk ^{\prime}-\vk).\vr_i}
\lbrack \bd_{i\sigma ^{\prime}}b_{i\sigma} + 
\delta _{\sigma \sigma ^{\prime}}h^{\dag}_ih_i \rbrack, $$  
Notice that right-hand side, which arises from the 
on-site anticommmutators, consists of Fourier components
of charge and spin-density operators. Suppose, as an approximation,
we replace the right-hand side by its average value and assume 
that there is no magnetic or CDW order. Then, we obtain  
$$ \lbrace c_{\vk\sigma},\cd _{\vk^{\prime},\sigma^{\prime}}\rbrace
\approx \xi\delta _{\sigma,\sigma ^{\prime}}\delta _{\vk\vk ^{\prime}},
$$
where $\xi = (1 + \delta)/2$. In other words, $c_{\vk\sigma}$ is
effectively replaced by $\xi ^{\half}$ times an usual fermion
operator which carries spin. If $k_F$ is chosen to yield the correct 
electron density in the same approximation, then the new Fermi 
surface agrees with the modified one, and the energy is given 
by $E_{bound}$. Since spin- and charge-density fluctuations are
completely suppressed, this state does not have any localized 
moment, and hence, there is no spin-charge separation.             

\bigskip

\noin IV. KINETIC ENERGY OF THE MEAN-FIELD STATES

\bigskip

The kinetic energy of an actual state is greater than 
$E_{bound}$. Now, the momentum distribution for the MF states 
is given by a convolution: 
\beq n_{c}(\vk) =  \frac{1}{N}\sum _{\vq} 
n_b(\vk + \vq)(1 \pm n_h(\vq)), \eeq
where $n_b(\vk)$ and $n_h(\vk)$ are the momentum distribution
functions for spinons and holons, respectively. The minus (plus)
sign corresponds to the Schwinger (slave) boson representation.
Consider the MF states at $T = 0$.

a) {\em Slave-boson State}: We consider an RVB type state in
which bosonic holons condense at $\vk = 0$,  giving
\beq n_c(\vk) = 1 - \delta (1 - n_b(\vk)).\eeq
Spinons form their own Fermi sea, identical to that for noninteracting
electrons, and $n_b(\vk) = 2$ for occupied and $0$ for
unoccupied states. Hence, electrons have the 
same Fermi surface, but $n_c(\vk)$ equals  $1 + \delta$ below, 
and $1 - \delta$ for all $\vk$ above the Fermi level all the way
to the top of the band. Since $\sum _{\vk} \epsilon (\vk) 
= 0$, the kinetic energy per site is given by
\beq E_{slave} = \delta E_{free}(1-\delta). \eeq

b) {\em Schwinger-boson State}: We consider the 
ferromagnetic state (small $J$) since the kinetic energy is
lowest in this case. (At larger $J$, the kinetic
energy, as defined here, is larger due to the 
competition with antiferromagnetic fluctuations).
Spinons condense at $\vk = 0$, so that
\beq n_c(\vk) = (1 - \delta) (1 - n_h(-\vk)). \eeq
The holons form a Fermi surface, centered at the zone corner, 
with $n_h(\vk) = 1$ in the occupied region, and zero in the
unoccupied region. The electrons have the same Fermi 
surface, so the electron Fermi energy is much higher than the  
noninteracting one. On the average, there are only $1 - \delta$ 
electrons below this Fermi level (even less than the number
at half filling!) and none above. In this case the kinetic energy 
per site is
\beq E_{sch} = \frac{(1-\delta)}{2}E_{free}(2\delta). \eeq

Fig (1) and (2) shows $n_c(\vk)$ and the kinetic energy for
various states. First, note that all states have the correct
kinetic energy ($=$ zero) at half filling. Away from half
filling, KE for either MF state is much larger than 
$E_{bound}$, by about a factor of two in the small-$\delta$ 
region. Surprisingly, for $\delta < 1/3$, the Schwinger boson 
state has a lower kinetic energy than the slave boson state.
One expects Schwinger boson theory 
to do better as far as exchange energy is concerned since
it works well for the Heisenberg model.
It appears that it is also better for the KE at low doping,
the region of interest for the cuprates. 
In the slave boson case, the Fermi surface coincides with
the noninteracting one, and below the Fermi level, $n_c(\vk)$ 
is the maximum ($ = 1 + \delta$) allowed by the sum rule.
However, above the Fermi level, $n_c(\vk)$ is large 
($= 1-\delta$) all the way to the top of the band, whereas,
it is zero for the other states. This
costs considerable amount of kinetic energy.
In the Schwinger boson case, the top of the
band is empty, which is fine, 
but the kinetic energy is still large compared
with $E_{bound}$ because the Fermi energy is much higher, 
and because $n_c(\vk) = 1-\delta$ in the occupied region,
smaller than value allowed by the
sum rule, all the way to the bottom of the band. 

c) {\em Hartree-Fock States in the Hubbard Model:}
How are these results affected by the constraints? That 
there is an error due to the average treatment of the
constraints can be seen from the average electron density
$$n_c = \sum _{\sigma}<\cd_{i\sigma}c_{i\sigma}> 
= \frac{1}{N}\sum _{\vk}n_c(\vk),$$
which does not equal its actual value $1 - \delta$ in
the MF theory. Instead, 
it equals $(1-\delta)^2$ in the Schwinger boson state,
and $(1-\delta^2)$ in the slave boson state. To assess the
role of the constraints we consider the HF states of the
Hubbard model in which constraints are treated exactly. 
We consider $U = \infty$ limit since  
mapping onto the $t-J$ model is exact (with $J = 0$). 
The HF ground state is feromagnetic,  
with the up-spin band filled up to the holon Fermi
level. The down-spin band is empty.  This is the
same ferromagnetic state found in SWB theory,
and has the same fermi surface. However, now 
$n_c(\vk) = n_{\uparrow}(\vk)+n_{\downarrow}(\vk) = 1$ 
in the occupied region, instead of $1 - \delta$. 
Since, as shown rigorously in section V., the states 
are essentially the same, the difference 
between the two results is a measure of the error (due
to the constraints) in the Schwinger-boson theory.

Since the Fermi levels for the HF and Schwinger
boson states coincide, the energies are proportional:
$$ E_{sch} = (1-\delta)E_{hub}. $$ Now, for all the states,
$E_{kin} < 0$ at finite $\delta$. Hence, $E_{hub}$ is less than
$E_{sch}$ (and also less than $E_{slave}$) for all $\delta$ 
(see Fig. 2).
At small $\delta$, $E_{hub}$ scales
with $\delta$, therefore the corrections due to constraints
are small, i.e., of order $\delta ^2$.  To put it another way,
since $E_{sch}/E_{hub} = 1 - \delta$,
the effect of the constraints in the physically important
small $\delta$ region is not large.

We see that kinetic energies of the MF and HF states are
much larger than $E_{bound}$, the lowest possible KE 
without double occupancy. In the small
$\delta$ region, $E_{bound}/E_{MF}$ is between 1.5
to 2, i.e., the difference is comparable to the
energy itself (Fig 3). This does not mean that $E_{bound}$ 
corresponds to the actual state of the system. The key point
is that $n_c(\vk)$ considerably smaller (by an amount 
$\propto \delta$ -- the scale of KE) than the allowed
value near the bottom of the band for the SWB and HF states,
and considerably larger near the top of the band for the 
slave boson states.  In the actual ground state, 
$n_c(\vk)$ is sure to depend on $\epsilon (\vk)$, and
is likely to be close to
$1 + \delta$ near the bottom, and close to zero near
the top of the band. This means that fluctuations
(in the MF states) that transfer electrons
from near the top of the band to near the bottom
will be dominant.     
     
\bigskip

\noin V. CONSTRAINTS: FROM SCHWINGER BOSONS TO SPIN WAVES

\bigskip

In this section we solve the constraint problem in the
$U = \infty$ limit. It is convenient to use the functional 
integral technique.
The Bose fields can be expressed as: $b_{i\sigma} = 
r_{\sigma}e^{i\phi _{i\sigma}}$. The constraint condition is
given by
$$r^2_{i\uparrow} + r^2_{i\downarrow} + h^*_ih_i = 1,$$ 
which does not depend on the phases. The action is 
invariant under the gauge transformation: 
$b_{i\sigma} \rightarrow   b_{i\sigma}e^{i\chi_i},~~~h_i \rightarrow
h_ie^{i\chi _i}.$ 
We can choose $\chi _i= - \phi _{i\uparrow}$, (or, equivalently
select a gauge:  $\phi _{i\uparrow} = 0$). 
The action is now independent of $\phi _{i\uparrow}$, which can be 
integrated out. The constraints are not effected. Next, we
introduce the constraints into the functional integrals as $\delta$
functions, and use the latter to integrate out $r_{i\uparrow}$.
This is equivalent to the replacement
$b_{i\uparrow} = r_{i\uparrow} = (1 - r^2_{i\downarrow} - 
h^{*}_ih_i)^{1/2}$ in the action. The remaining degrees of freedom 
define the generalized HP representation. It is easy to see 
that the corresponding Hamiltonian is obtained from the
Schwinger boson Hamiltonian by using the following substitutions:  
$$b_{i\downarrow} \rightarrow \cb_i, 
~~~ h_i \rightarrow \ch _i, $$
$$\bd _{i,\uparrow} = b_{i,\uparrow} \rightarrow 
(1 - \cbd_i\cb _i - \chd_i\ch_i )^{\half}. $$
The spin operators are represented as
$$S^{+}_i = (1 - \cbd_i\cb _i - \chd_i\ch_i )^{\half}\cb_i. $$
$$S^{z}_i = \half(1 - \chd_i\ch_i) - \cbd_i\cb _i. $$ 
In the absence of the holes, one recovers the usual HP representation 
for the Heisenberg ferromagnet. The generalization to the antiferromagnetic
case is straightforward.  As in the usual HP representation, the square-root 
represents an anholonomic constraint. This is of course not as stringent
as the constraints in the Schwinger-boson (or slave-boson) representation.
The electron operators are given by
$$\cd _{i\downarrow} = \cbd _i\ch _i $$
$$\cd _{i\uparrow} = (1 - \cbd_i\cb _i - \chd_i\ch_i )^{\half}\ch _i.$$
In the last expression, we can drop the hole density operator within 
the square-root since it will not contribute (because
$\ch$ is fermionic). This gives,
$\cd _{i\uparrow} = (1 - \cbd_i\cb _i)^{\half}\ch _i$.
Since the constraints have been integrated out, there is no
gauge freedom left. Note that the $\cb$ and $\ch$ are gauge  
invariant since they are proportional to gauge-invariant
operators $S^{+}$ and $\cd$, respectively.
As in the Heisenberg case, $\cbd$ creates a spin-wave 
excitation -- a magnon.  Similarly, the operator $\chd$  
creates a spinless hole excitation (not the unobservable holon 
of the SWB representation).  
Other difficulties associated with the 
constraints also disappear. For example, the local
density of electrons is given by
$$\sum _{i\sigma}\cd _{i\sigma}c_{i\sigma} = 1 - \chd_i\ch_i,$$ 
which correctly yields, $n_c = 1 - \delta$, in contrast to,
$n_c = (1-\delta)^2$ in the Schwinger boson MF theory.

For $J = 0$ ($U = \infty$) the $t$-$J$ Hamiltonian, in the
generalized HP representation becomes
\beq H = t\sum _{ij}\lbrack\lbrace (1-\cbd _i\cb _i)^{\half}
(1-\cbd _j\cb _j)^{\half} + \cbd _i\cb _j\rbrace\chd _j\ch _i 
+ h.c.\rbrack. \eeq  
Note that the total no of holes ($N_h$) and spin-wave
excitations ($N_b$) are 
separately conserved. This follows from the conservation of 
the total electron number, $N_c = N - N_h$, and total magnetization
$M_z = \sum _iS^{z}_i = \half (N - N_h) -N_b$
in the parent Hubbard model. 
We will keep $N_c$ fixed, which fixes $N_h$. But $N_b$ is not fixed
in the ordered state, but varies from $0$ to $N_c$, where 
$N_b = N_c$ corresponds to zero magnetization. 

We can diagonalize the Hamiltonian separately in each $N_b$ 
subspace. The $N_b = 0$ subspace only has holes, 
and is of particular interest. In this case, $\ch _i = 
\cd_{i\uparrow}$. All electrons have spin up, the average 
magnetization, $m = (1 - \delta)/2$, has the largest possible value.
Within this subspace, the Hamiltonian reduces to
\beq H_h = t\sum _{ij}(\chd _i\ch _j + h.c), \eeq
with no residual interaction. Note the
this is nothing but the $U = \infty$ Hubbard Hamiltonian  
in the Hartree-Fock approximation in the fully spin-polarized
subspace. In particular, it contains the ferromagnetic HF
ground state (the generalized Nagaoka state) considered previously.
We have thus shown that HF theory can be obtained from the
Schwinger boson theory, and it takes the constraints into account
exactly.

All wave functions corresponding to $H_h$ are one-component
Slater determinants, describing the charge sector. These are also exact 
wave functions of the full Hamiltonian (actually for all $U$) since in the 
$N_b = 0$ subspace, $H_h$ is the exact Hamiltonian. It is not known 
rigorously whether the Nagaoka state is the true ground state at finite 
$\delta$.  A possible answer is that the true ground state belongs to 
a different $N_b$ subspace, with magnitization  $m = 
\half(1 -\delta) - n_b$, which decreases continuously to zero 
$\delta _{cr}$. Here, $n_b = N_b/N$ is the magnon density.

For finite $n_b$, we need to include the magnons. For small $n_b$,
we expand the square root to leading order in $n_b$, which gives 
\beq H =  t\sum _{ij}(\chd_i\ch_j + h.c) + t\sum _{ij} 
\lbrack(\cbd _j\cb _i - \half \cbd _i\cb _i - 
\half\cbd _j\cb _j)\chd_i\ch_j + h. c. \rbrack.\eeq 
The second term includes spin-wave effects, and extends the
treatment of the $U = \infty$ Hubbard model beyond the
Hartree-Fock level.

Note that there is no free magnon terms in the Hamiltonian.
This is very different from the Heisenberg ferromagnet, where
the lowest order ($1/S$) term is quadratic in bosonic operators.
Here, magnons (spin)  interact with holes (charge) at the lowest 
level. The simplest approximation is to do a MF decomposition
of the quartic term, leading to quadratic Hamiltonians for holes 
and magnons: $H = H_b + H_h$.   
The hole Hamiltonian has the HF form (Eq. 23), but with a
renormalized hopping parameter 
$$t_h = t(1 + B - n_b), $$
where, $B = <\cbd_i\cb_j>$. The bosonic mean-field Hamiltonian
is given by
\beq H_b = tD\sum _{ij}\lbrack \cbd _i\cb _j + \cbd _j\cb _i
- \cbd _i\cb _i - \cbd _j\cb _j\rbrack, \eeq
where $D = <\chd _i\ch _j>$. The magnon Hamiltonian has the same
form as the one in the Heisenberg model. The magnon energy is then
given by,
\beq \omega _m(\vk) = 2tD(\cos k_x + \cos k_y - 2). \eeq
It is easily shown that $D \le 0$ is negative, so that
$\omega _m(\vk) \ge 0$. Therefore, there are no magnons at 
$T = 0$, and the Nagaoka state found in the HF approximation 
continues to be the ground state at this level of approximation.  
At finite $T$, magnons will proliferate just as in Heisenberg 
ferromagnet, and destroy long-range order.
This is the correct result (Marmin-Wagner theorem).
In principle, Takahashi's modified spin-wave method \cite{tak} can 
be used to describe the system at finite $T$.   

\medskip 

\noin {\em Electron Green's Function and Missing Spectral Weight}:

\medskip 

To obtain the Green's function for the up electrons, we expand the
square-root. Now, at the MF level,
$$T_{\tau}<c _{i\uparrow}(\tau)\cd_{j\uparrow}(\tau ^{\prime}) >
= T_{\tau}<(1 - \half \cbd _i(\tau)\cb _i(\tau) - 
\half \cbd _j(\tau ^{\prime})\cb _j(\tau ^{\prime}))>  
<\chd _i(\tau)\ch _j(\tau ^{\prime})>. $$
The average of the density operators simply gives the magnon density
$n_b$, which is uniform. Hence, the Green's function is proportional
to the hole Green's fumction: 
\beq G_{c\uparrow}(\vk,\omega) = \frac{1 - n_b}{i\omega + 
\epsilon _{\ch}(\vk)}, \eeq
where $\epsilon _h(\vk) = 2t_h(\cos k_x + \cos k_y)- \mu _h$
is the hole spectrum, which includes the chemical potential
$\mu _h$. This term is analogous to the condensate part
in the Schwinger boson MF theory \cite{sar4}.

The Green's function for down electrons is a convolution of magnon
and hole Green's functions, and is given by \cite{ark}
\beq G_{c\downarrow}(\vk,\omega) = \frac{1}{N}\sum _{\vq}
\frac{f(\epsilon _{\ch}(\vq)) + 
n(\omega _m(\vk + \vq))}{i\omega - \omega _m(\vk+\vq)+\epsilon _{\ch}(\vq)}.
\eeq
This term is the analog of the non-condensate part in the SWB theory.
The corresponding spectral function is nonzero only for $\omega > 0$ 
and is incoherent.
It is absent in the HF theory. In the HF case there are two bands,
described by simple poles. The up-spin band is the
lower Hubbard band. The down-spin band is the upper
Hubbard band which  represents the doubly occupied sector 
amd is separated by $U \rightarrow \infty$; it is
is absent in our case. In the simple HF theory,
the spectral weights for each band is unity. This does not
agree with the sum rule, which gives $1 + \delta$ and 
$1 - \delta$, respectively. The reason for the discrepancy 
is that magnons are missing in the simple HF theory. The 
expression for $G_{\downarrow}$ (Eq. 28), is precisely this 
(low-frequency) magnon 
contribution, with a total spectral weight of $n_b + \delta$. 
Together with the spectral weight of $G_{\uparrow}$, this gives a
total spectral weight of $1 + \delta$, in agreement with the
sum rule.

The Green's function in the spin-wave theory is very similar to that
in the Schwinger boson theory \cite{sar4}. However, in the latter  
case, magnetic ordering is along the $x$ direction, so that
the Green's function for up and down electrons are identical. 
For a proper comparison we need to take the average: 
$G_c = (G_{c\uparrow}+G_{c\downarrow})/2$. Although the structure 
of $G_c$ in the two cases is the same, the spectral
weight distribution between the coherent (pole) and 
incoherent pieces is different.
For a comparison, it is convenient to express these in terms of 
the magnetization $m$ which is the same in both theories, and equals
$(1-\delta)$ at $T = 0$. Recall that $n_c(\vk)$, and hence, the kinetic
energy, is determined by the negative frequency part, i.e., by  
the pole piece ($G_{c\uparrow}$), which carries
a spectral weight = $\half(1 + \delta) + m$, which is greater than the
(total) weight in the SB theory $ = 2m$.  In particular, it reduces to
the HF value ($= 1$)at $T = 0$. Therefore, HF theory gives the best kinetic 
energy.

\bigskip 

\noin VI. CONCLUSIONS

\bigskip

In this paper we have shown that mean-field states of the $t$-$J$
model exhibit complete spin-charge separation in the sense of
Eq. (2), and in $d = 2$, have high kinetic energies.   
This supports qualitative arguments, given
in previous papers, that spin-charge separation localizes electrons 
which costs kinetic energy. 
The determination of $E_{bound}$ is a key result since 
it includes excess KE which arises simply from the removal 
of double occupancy,
but not necessarily from spin-charge separation. 
The calculation of $n_c(\vk)$ is also interesting, because it 
shows that mean-field values differ substantially from what is 
allowed by the sum rule at the band edges. Therefore 
fluctuations that transfer electrons from the occupied to
the unoccupied regions, can, in principle, lower the energy 
significantly, i.e, by an amount comparable to KE itself.

The sum rule and the expression for $E_{bound}$
are valid for arbitrary $J$. The kinetic energies
presented here are for $U = \infty$ ($J = 0$) for which
KE for MF and HF theories can be compared directly. 
Extension to finite $J$ is 
not difficult since the MF problem has been
solved exactly, but numerically. At  finite $J$, the
KE of the $t$-$J$ model would be larger since
the two terms compete.  But conclusions
of this paper are not going to change. Of course, $J$ will
be important to describe the spin singlets and for 
superconductivity.

The second important result is the mapping of the $t$-$J$
model onto a generalized Holstein-Primakoff model. This
allowed us to solve the difficult problem of the
constraints for the SWB MF state at $U = \infty$. Our results
show that constraints do not change the nature of the state
-- it remains fully spin-charge separated and has the same magnetic
ordering. Extending the method to finite $J$ is somewhat
more complcated, but can be done. It will be useful to study
fluctuations such as spin-charge recombination within 
the generalized HP model so as to take into account the
constraints at the outset. 

The mapping also establishes the cannection between the SWB
theory and the HF approximation. The latter has long been
used to study magnetic ordering in strongly correlated 
systems. However, HF approximation fails at finite $T$
(or, at finite frequency) because spin-waves are not included.
For example, $T_c$ at which magnetic order vanishes
scales as $U$. The inclusion of the spin-waves via
the mapping solves this problem, as spin-waves destroy  
LRO in 2-d at any finite $T$, and in $d = 3$ at a temperature 
$\sim t$ (as $U \rightarrow \infty$). It also demonstrates
complete spin-charge separation in the HF plus spin-wave theory.
This method will be useful in studying itinerant ferromagnetism
in  $d = 3$.

Our results suggest that fluctuations that result in 
spin-charge recombination are likely to be strong. 
Such processes lower KE by moving electrons from above the 
Fermi level to the bottom of the band. In previous papers
we have treated recombination by RPA in which the
physical electron appears as a fermionic collective 
mode \cite{sar1}. In $d = 2$, it helps destroy the magnetic LRO,
but, for $\delta$ not too large, the Fermi liquid state
is not recovered. For finite $J$, additional spin-charge
recombination between a pair of holons with preformed pair
of spinons leads to $d$-wave superconductivity \cite{sar2}.

For the $U = \infty$ case, the destabilization of the HF
state has been demonstrated by Shastry, Krishnamurthy and 
Anderson \cite{sha}. They considered the destruction of  
ferromagnetic LRO away from half-filling.  Consider
the ferromagnetic ground state with all spin up. 
Since $n_{k\uparrow} = 1$ in the occupied region,
the only way kinetic energy can be lowered is by destroying
an up spin electron at the Fermi level, and creating
a down-spin electron at the bottom of the band. But 
creating a bare down-spin electron would cost $U$. Instead
one must therefore create a projected down-spin electron.
They showed that the MF 
state becomes unstable with respect to such spin modes.
Note that in HP representation, (or SWB representation),
the projected electron is a collective object
created by $\cd _{\vk\sigma} = 
N^{-1/2}\sum _{\vq}\cbd_{\vk+\vq,\sigma}h_{\vq}$.
This process therefore corresponds to spin-charge binding,
and is the same one considered by us.

\bigskip

\centerline{\bf FIGURE CAPTIONS}

\noindent {\bf Fig.1:}

\medskip

Momentum distribution for various states as a function of
energy. The slave boson case has a high kinetic energy 
because $n_c$ is large all the way to the top of the band. 
For the other three states the top of the band is empty, but
for the Schwinger boson case the energy is again large
because $n_c$ is small at the bottom of the band.    

\medskip
 
\noindent {\bf Fig.2:}

\medskip

Kinetic energy as a function of hole density $\delta$. The
HF energy is lower than either Schwinger boson or slave-boson
case. However, $E_{bound}$ is much lower than $E_{bound}$.
The discrepancy between the HF and Schwinger boson energies
measures the error due to poor treatment of constraints.     

\medskip 
\noindent {\bf Fig.3:}

\medskip

$E_{bound}$/$E_{hub}$ as a function of hole density. 
At small $\delta$, the HF energy is larger by about
a factor of two. (Note both energies are negative).


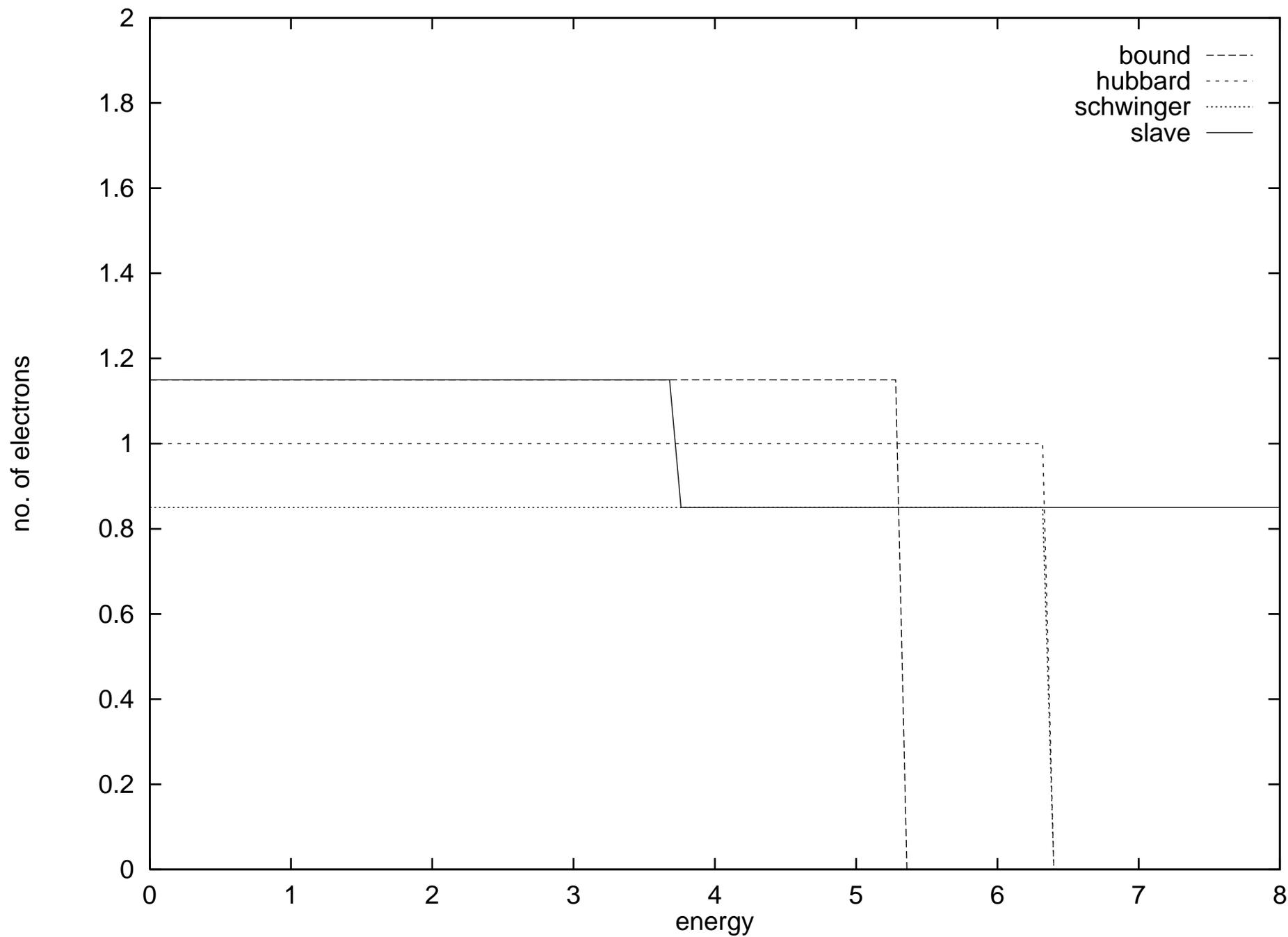
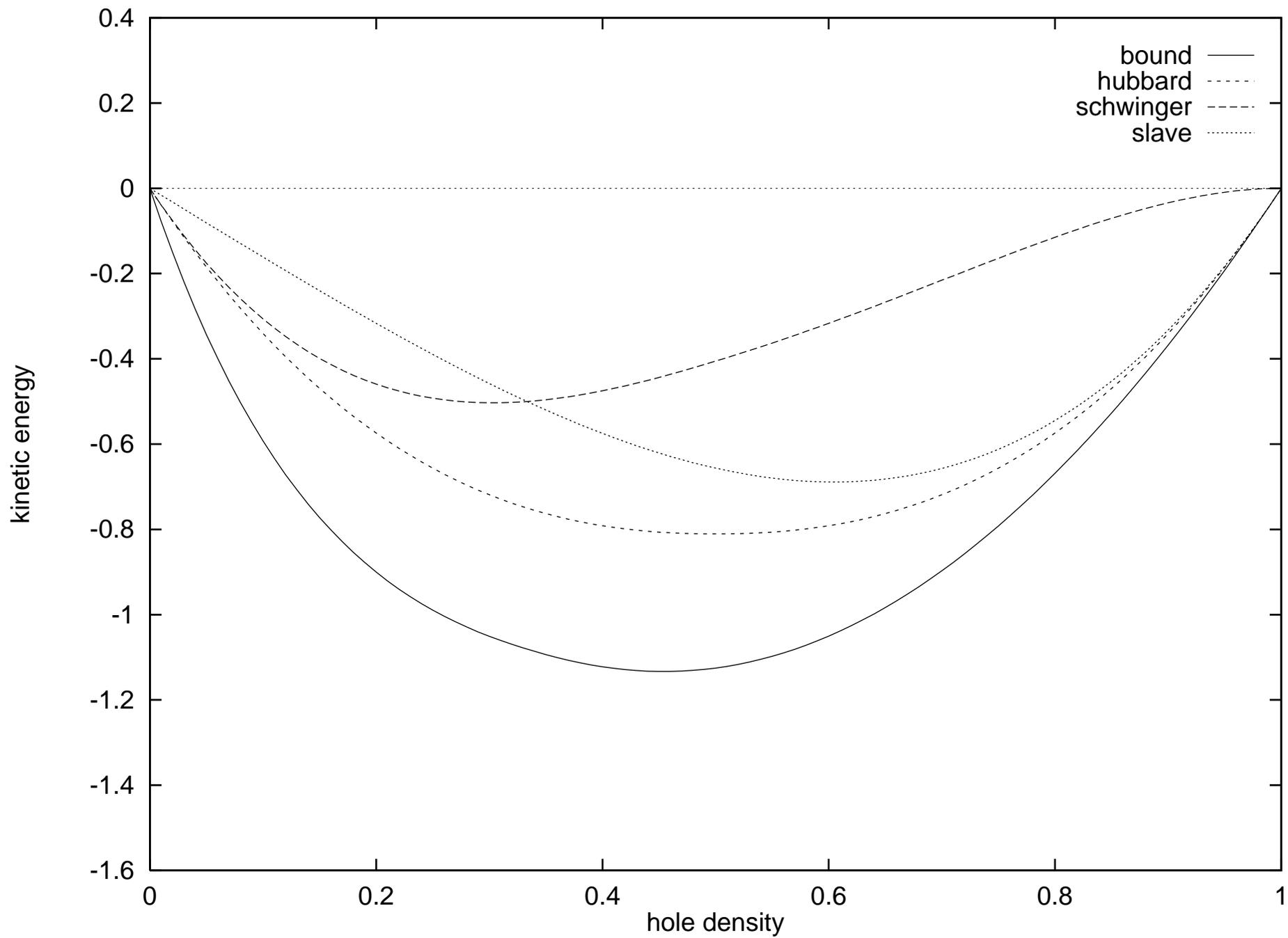
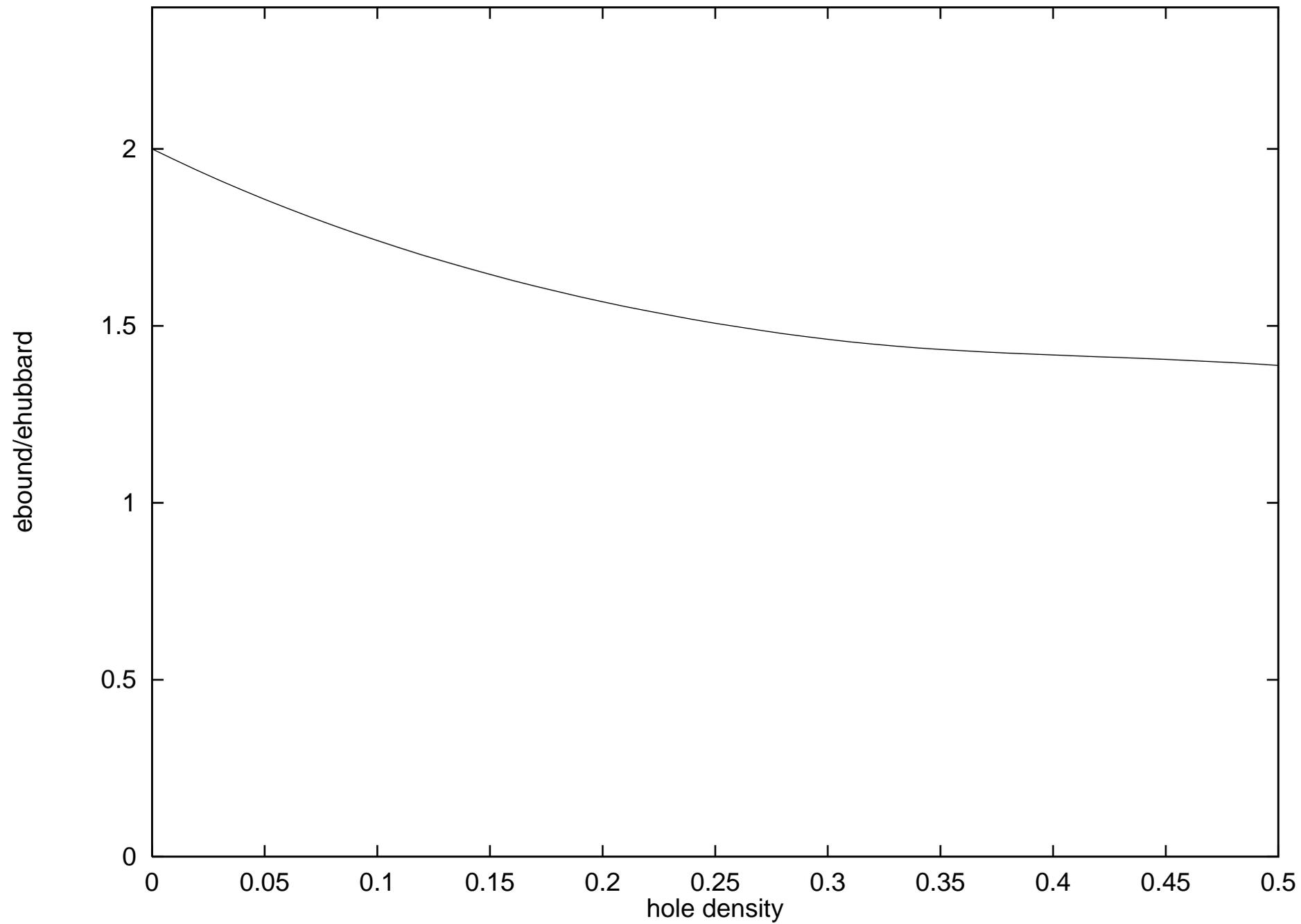
\end{document}